\documentclass[conference]{IEEEtran}
\IEEEoverridecommandlockouts
\usepackage{cite}
\usepackage{amsmath,amssymb,amsfonts}
\usepackage{graphicx, enumitem}
\usepackage{textcomp}
\usepackage[ruled,vlined,linesnumbered]{algorithm2e}
\usepackage[table,dvipsnames]{xcolor}
\usepackage{tikz}
\usetikzlibrary{shapes.geometric, arrows, quantikz, backgrounds}
\usepackage{svg}
\usepackage{bm}
\usepackage{makecell}
\usepackage{booktabs}
\usepackage{multirow}
\usepackage{url}
\usepackage{tikz}
\usepackage{soul}
\usepackage[none]{hyphenat}
\usepackage{subfigure}
\usepackage{hyperref}
\usepackage{braket}
\usepackage{adjustbox}
\SetKwIF{If}{ElseIf}{Else}{if}{}{else if}{else}{end if}%
\SetKwInOut{Parameter}{Params}

\newcommand{\var}{\texttt}

\newcolumntype{L}[1]{>{\raggedright\let\newline\\\arraybackslash\hspace{0pt}}m{#1}}
\newcolumntype{C}[1]{>{\centering\let\newline\\\arraybackslash\hspace{0pt}}m{#1}}
\newcolumntype{R}[1]{>{\raggedleft\let\newline\\\arraybackslash\hspace{0pt}}m{#1}} 

\def\BibTeX{{\rm B\kern-.05em{\sc i\kern-.025em b}\kern-.08em
    T\kern-.1667em\lower.7ex\hbox{E}\kern-.125emX}}

\tikzstyle{startstop} = [rectangle, rounded corners, minimum width=2cm, minimum height=1cm,text centered, draw=black]
\tikzstyle{process} = [rectangle, minimum width=2cm, minimum height=1cm, text centered, draw=black]
\tikzstyle{decision} = [diamond, minimum width=2cm, minimum height=1cm, text centered, draw=black, aspect=2]
\tikzstyle{arrow} = [thick,->,>=stealth]
\tikzstyle{arrowR} = [thick,->,>=stealth, dashed]
\tikzstyle{coord} = [coordinate]

\begin{document}
	
\title{Hybrid Quantum-Classical Heuristic\\for the Bin Packing Problem}

\author{
\IEEEauthorblockN{
    Mikel Garcia de Andoin \IEEEauthorrefmark{3}\IEEEauthorrefmark{2}\IEEEauthorrefmark{6}\IEEEauthorrefmark{1}
    Eneko Osaba\IEEEauthorrefmark{2}\IEEEauthorrefmark{1},
	Izaskun Oregi\IEEEauthorrefmark{2},
	Esther Villar-Rodriguez\IEEEauthorrefmark{2},
	Mikel Sanz\IEEEauthorrefmark{3}\IEEEauthorrefmark{6}\IEEEauthorrefmark{4}\IEEEauthorrefmark{5},
	}
	\IEEEauthorblockA{\IEEEauthorrefmark{3}Department of Physical Chemistry, University of the Basque Country UPV/EHU, Apartado 644, 48940 Leioa, Spain}
	\IEEEauthorblockA{\IEEEauthorrefmark{2}TECNALIA, Basque Research and Technology Alliance (BRTA), 48160 Derio, Spain}
	\IEEEauthorblockA{\IEEEauthorrefmark{6}EHU Quantum Center, University of the Basque Country UPV/EHU, 48940 Leioa, Spain}
	\IEEEauthorblockA{\IEEEauthorrefmark{4}IKERBASQUE, Basque Foundation for Science, Plaza Euskadi 5, 48009 Bilbao, Spain}
	\IEEEauthorblockA{\IEEEauthorrefmark{5}BCAM - Basque Center for Applied Mathematics, Alameda de Mazarredo 14, 48009 Bilbao, Spain}
	\IEEEauthorblockA{\IEEEauthorrefmark{1}Corresponding authors. Emails: mikel.garciadeandoin@ehu.eus, eneko.osaba@tecnalia.com}}
	
\maketitle

\begin{abstract}
Optimization problems is one of the most challenging applications of quantum computers, as well as one of the most relevants. As a consequence, it has attracted huge efforts to obtain a speedup over classical algorithms using quantum resources. Up to now, many problems of different nature have been addressed through the perspective of this revolutionary computation paradigm, but there are still many open questions. In this work, a hybrid classical-quantum approach is presented for dealing with the one-dimensional Bin Packing Problem (1dBPP). The algorithm comprises two modules, each one designed for being executed in different computational ecosystems. First, a quantum subroutine seeks a set of feasible bin configurations of the problem at hand. Secondly, a classical computation subroutine builds complete solutions to the problem from the subsets given by the quantum subroutine. Being a hybrid solver, we have called our method H-BPP. To test our algorithm, we have built 18 different 1dBPP instances as a benchmarking set, in which we analyse the fitness, the number of solutions and the performance of the QC subroutine. Based on these figures of merit we verify that H-BPP is a valid technique to address the 1dBPP.
\end{abstract}


\section{Introduction} \label{sec:intro}

Computational optimization is a highly studied research topic within the wider Artificial Intelligence field. Its contrasted applicability is one of the main reasons of the success of this knowledge area, which is resorted to address a myriad of real-world oriented tasks.

Usually, the efficient tackling of optimization problems involves the need of significant computational resources, making impractical the adoption of a brute-force approach. For this reason, the formulation of time-efficient solving strategies has emerged as a hot topic, leading to the design of a plethora of heterogeneous techniques during the last decades. 

At the time this paper is written, the vast majority of optimization algorithms have been conceived for being executed on classical computers. In this context, Quantum Computing (QC, \cite{steane1998quantum}) has recently emerged as a promising paradigm for facing optimization problems. In a nutshell, QC provides a revolutionary approach for dealing with complex problems with a possible significant advantage\cite{ajagekar2020quantum}. Today, QC is the pivotal point of a growing amount of experimental and theoretical scientific studies \cite{navaneeth2021study,saki2021survey,herman2022survey}.

Despite the potential demonstrated by QC, this research field is still at its dawn, and the available quantum processors present some limitations in terms of performance and capability \cite{fellous2020limitations,al2017natural}. Issues such as the reduced size of the quantum processors, the inaccurate control of the quantum resources, or the limitations in material science should be faced by researchers and engineers on their effort towards fault-tolerant quantum-computing. As a result of these handicaps, two kind of solvers prevail in the current literature: i) \textit{purely quantum} approaches, whose objective is to face a problem employing only QC resources, and \textit{quantum-classical hybrid} techniques, which are conceived to enhance applicability within the short and mid-term QC limitations. 

In this context, our objective with this paper is to take a step forward in QC by proposing a \textit{quantum-classical hybrid} algorithm for addressing one of the best-known combinatorial optimization problems: the Bin Packing Problem (BPP, \cite{garey1981approximation}). Although it is a canonical use case, the BPP is often emerges in a wide variety of industrial problems. To our best knowledge, this problem has not yet been faced from a QC perspective. Thus, the present work represents a step further over the current QC literature, presenting the following contributions:

\begin{itemize}
	\item We introduce a \textit{quantum-classical hybrid} algorithm for solving the well-known one-dimensional BPP (1dBPP,\cite{munien2020metaheuristic}). The algorithm is composed of two modules, each one designed for being executed in different computational ecosystems. First, a QC-based module is employed for finding the set of feasible bin configurations of the problem at hand. In other words, this module seeks as many feasible bins as possible, considering the constraints of the problem (i.e. the weight of the items and the capacity of the bins available). The second module, which is implemented to be run in classical computers, takes as input the list of feasible subsets provided by the QC-based module, and generates complete solutions to the problem finding the optimal combination of subsets. As it is a hybrid solver, we have called our method H-BPP. To the best of our knowledge, it is the first time that a BPP is solved using such strategy. 
	
	\item As mentioned, the proposed method has been developed for facing the 1dBPP. Despite related problems, such as the knapsack problem \cite{pusey2020adiabatic} or the subset-sum problem \cite{bonnetain2020improved} have been considered in the literature, no variant of the BPP has been studied from a QC perspective up to now. Therefore, taking into account the scientific and business interest that BPP-related problems still arise in the community and in the industry \cite{aydin2019bin,aydin2020multi}, it is relevant to further research in this direction. For testing the performance and accuracy of our method, we have conducted a preliminary but extensive experimentation over 18 synthetically-generated heterogeneous instances of the 1dBPP, with sizes ranging from 10 to 12 items. Results obtained by our algorithm are compared with the ones obtained by a brute-force algorithm.
	
\end{itemize}

The rest of this manuscript is structured as follows: we present in Section \ref{sec:back} a brief introduction on QC and BPP. Afterwards, the problem addressed in this paper, the 1dBPP, is described in Section \ref{sec:BPP}. In Section \ref{sec:solver}, we delve into the main characteristics of our hybrid algorithm. Experimental setup and results are discussed in Section \ref{sec:exp}. Finally, in Section \ref{sec:conc} we summarize our conclusions and outline future research lines.

\section{Background} \label{sec:back}

This section briefly introduces the two main pillars of this paper: Quantum Computing, in Section \ref{sec:QCBack}; and the Bin Packing Problem, in Section \ref{sec:BPPBack}.

\subsection{Introduction to Quantum Computing}\label{sec:QCBack}

Quantum computing takes advantage of quantum resources to solve computational problems. Indeed, exploiting  entanglement and superposition properties, some quantum algorithms have shown potential to speed up the convergence towards a solution for certain problems. Two representative algorithms that show quantum advantage are Grover's algorithm for the unstructured searching problem \cite{grover1996search} and Shor's algorithm for the integer factorization \cite{shor1994factoring}. Even though in these two cases a speedup is achieved, not every problem can benefit from employing a quantum resources \cite{scott2010BQP}, especially in the current Near-Intermediate Scale Quantum era \cite{frank2020bittertruth, daniel2021limitationsNISQ} (NISQ, noisy quantum hardware with a size in the order of 10-100 qubits \cite{Preskill2018NISQ}). However, quantum algorithms might have a speedup compared with classical approximate algorithms for particularly hard problems (i.e. problems in the approxNP-hard set), such as the Max-E3-SAT problem \cite{johan2001MAXE3SAT}, \cite{younes2015quantMAXE3SAT}.

Within the framework of quantum computing, there are three main different paradigms. Firstly, digital quantum computing (DQC) \cite{Deutsch1989DQC}, which works with a sequence of ideal operations (or gates) taken from an universal set. For implementing algorithms in this paradigms, we have a pletora of gate-based quantum computers (IBM\footnote{\url{https://www.ibm.com/quantum-computing/}}, Google\footnote{\url{https://www.quantumai.google/}}, Rigetti\footnote{\url{https://www.rigetti.com/}}, ColdQuanta\footnote{\url{https://www.coldquanta.com/}}), which allows users to run programs written as the application of seccessive gates. Then, analog quantum computing (AQC) is a less flexible but much more robust approach \cite{johnson2014AQC}. In AQC, a controllable quantum system is manipulated to mimic the system of interest. This approach also includes quantum annealing and simulators, which can be implemenmted in any system which provides a way to adiabatically modify the system Hamiltonian (D-Wave\footnote{\url{https://www.dwavesys.com/}}, Xanadu\footnote{\url{https://www.xanadu.ai/}}). finally, an universal quantum computing paradigm has been proposed, namely digital-analog (DAQC \cite{adrian2020DAQC}). This approach combines the robustness of AQC with the flexibility of DQC. The the algorithms are implemented by alternating the application of (digital) single-qubit gates with analog entangling blocks. Running a circuit using DAQC requires a hardware in which one has access to fast single-qubit gates and an interaction Hamiltonian, as it has been implemented in different platforms.

Encouraged by early results, the quantum-computing and computer science-communities have conducted a huge effort for achieving quantum advantage. As a prominent example of this research, an information processing task performed by a QC was proven unaffordable for any classical computer. Even though this has been attained for certain artificial problems \cite{Arute2019GoogleSupremacy, JianWei2021Advantage}, achieving this for a useful application is still an open question. Approximate optimization problems are considered a suitable target for this goal. On this regard, quantum annealing takes advantage of a purely quantum mechanism, the adiabatic theorem \cite{Born1928Tadiabatico, Kato1950Tadiabatico}, to find the ground state of a Hamiltonian which encodes the solution to the problem \cite{domino2021QArailway, hegade2021QAportfolio}. A family of near-term quantum algorithms that have shown a reasonable success are variational quantum algorithms (VQAs) \cite{Cerezo2021VQA}, with VQE and QAOA \cite{farhi2014QAOA} as the most prominent examples. These algorithms relay on the classical optimization of control parameters with the objective of obtaining the best possible solution given a certain amount of resources. On another vein, hybrid quantum algorithms employs quantum algorithms to speedup certain operations, while subrogating some tasks to classical routines \cite{Marsh2019qwalk, Ding2021LND}. When evaluating quantum algorithms, it should be noted that the probability to obtain a solution is unlikely to reach exactly $1$. On top of this, the problems of efficiently encoding information into a quantum system and retrieving it by means of measurements are generally open problems.

\subsection{BPP state of the art}\label{sec:BPPBack}

A paradigmatic example of a NP-Hard problem is the BPP, which is a classical combinatorial optimization problem whose objective is to pack a finite set of items into a group of available bins. More precisely, the goal is to minimize the total number of bins used not exceeding the fixed maximum capacity of each bin considered in the solution. The packaging of items or packages in different containers or bins is a daily and crucial task in the field of production and logistics. For this reason, multiple packaging problems can be formulated depending on the size of the items to be packed, as well as the size and capacity of the containers. These types of issues have been widely discussed in the literature for several decades, giving rise to remarkable survey and review papers \cite{man1996approximation,lodi2002two,lodi2002heuristic}. 

The 1dBPP is considered to be the simplest packaging problem. Anyway, despite its simplicity, it is applied to a wide variety of real-world problems \cite{van2007improving,angelelli2014optimal}. Furthermore, it is frequently used as a benchmarking problem \cite{fleszar2002new,ozcan2016novel}. Recent works focused on 1dBPP led to the proposition of novel methods for its resolutions, such as the cooperative parallel grouping genetic algorithm introduced in \cite{kucukyilmaz2018cooperative}, the branch-and-price-and-cut technique proposed in \cite{wei2020new}, or the adaptive fitness-dependent optimizer developed in \cite{abdul2020adaptive}. 

Beyond its basic formulation, several variants of the 1dBPP have been proposed in order to give an answer to some specific constraints. Some examples of these variants can be found in \cite{krause1975analysis}, in which a maximum number of items per bin is fixed; or in \cite{coffman1983dynamic}, in which items should be placed in bins according to some concrete time intervals. Further variants of the basic 1dBPP can be found in \cite{csirik2018variants}. In this present research, the canonical version of the 1dBPP is considered. This variant still gathers lot of attention from the community, as can be seen in recent surveys such as \cite{munien2021metaheuristic} and \cite{munien2020metaheuristic}.

Additionally, based on the original formulation of the 1dBPP, many variants have been recently proposed in recent literature to address real-world situations. On this regard, two are the most frequently referred variants of the BPP. The first one is the two dimensional BPP (2dBPP, \cite{lodi2014two}), in which each item is described by two different parameters: height and width. There are many works in the recent literature that revolve around this problem and its resolution. In \cite{laabadi2019crow}, for example, the authors address this problem with a genetic algorithm, adding a search mechanism called Crow Search to increase the exploration capacity of the algorithm. In the work \cite{lopez2019introducing} a hybrid approach applied to the same problem is presented. Additional examples can be found in articles like \cite{bezerra2020models} or \cite{sbai2019hybrid}. A second variant is the three-dimensional BPP (3dBPP, \cite{martello2000three}), in which each packet has three dimensions: height, width, and depth. Consequently, the containers in which the packages are stored have three dimensions as well. This variant has also been widely investigated by the community \cite{oliveira2019adaptive, pugliese2019solving}.

Furthermore, variations with a variable number of additional restrictions or constraints arising from real-world use cases have been studied, usually leading to an increment of the problem complexity. Some examples are \textit{i}) the minimum clearance of the container \cite{almeida2016resolution}, \textit{ii}) load balancing inside the container \cite{trivella2016load}, \textit{iii}) expiration dates of rapidly perishable products \cite{polyakovskiy2018hybrid}, \textit{iv}) transport restrictions \cite{paquay2018tailored}, or \textit{v}) possible packet fragmentation \cite{bertazzi2019bin}.

In addition to these restrictions, the problem has been treated from different computational approaches, highlighting the single-objective and multi-objective variants \cite{spencer2019greedy}. 

To sum up, the research around BPP-related problems is still vibrant today. Even the most basic variant of the BPP, the 1dBPP, still gathers a significant attention from the community. However, as far as we know, there is no work in the literature that deals with this problem from the QC perspective. On the contrary, there are few publications mentioning the BPP and quantum algorithms, but most of them discuss it from quantum-inspired classical algorithms perspective \cite{layeb2012cuckooquantum, toru2021quarterBPP, swain2014qaBPP}. Furthermore, slightly related with the 1dBPP, there are quantum algorithms solving the subset-sum problem \cite{bernstein2013subset, bonnetain2020improvedsubset, allcock2021dynamicalsubset}. However, these algorithms take advantage of some-problem specific properties which can not straightforwardly be translated to our scheme.

For this reason, this paper implies a remarkable step forward in the literature, tackling a BPP problem through a hybrid classical-quantum algorithm.

\section{The One Dimensional Bin Packing Problem}\label{sec:BPP}

As mentioned beforehand, this paper is focused on solving a BPP variant known as 1dBPP. The problem can be formally defined as follows \cite{munien2021metaheuristic}: considering a maximum amount of $n$ bins $\{B_1,B_2,\dots, B_n\}$ of equal non-negative capacity $C$, and a set of $n$ items $\mathcal{W}=\{w_1,w_2,\dots,w_n\}$, where $w_i$ is a positive value $0<w_i \le C$ representing the weight of item $i$, the objective is to find the smallest amount of bins $b\leq n$ such that partition $L=B_1 \cup B_2\ldots \cup B_b$ fits the whole $\mathcal{W}$. Furthermore, the sum of the item weights in each bin must not exceed $C$. 

Regarding the codification used for representing a solution to the problem, the well-known binary encoding has been considered. In this way, the 1bBPP can be formulated in the following way:

\begin{equation}
\min {b}  
\end{equation}
Subject to:
\begin{equation}\label{1dBPP_const1}
\sum_{i=1}^{n} w_{i}  x_i^{(j)} \leq C,\ \forall j \in \{1, \dots, b\},
\end{equation}
\begin{equation}\label{1dBPP_const2}
\sum_{j=1}^{b} x_{i}^{(j)} = 1,\ \forall i  \in \{1, \dots, n\},\\
\end{equation}
\begin{equation}\label{1dBPP_const4}
x_i^{(j)} \in \{0,1\},\ \forall i\in \{1, \dots, n\},\ \forall j\in\{1,\dots,b\},\\
\end{equation}
where $b$ is the number of employed bins, such that the set of bins is $\{B_1,...,B_b\}$, and $x_i^{(j)}=1$ if item $i$ is in bin $j$.

This strategy has the advantage of being fully compliant with digital QC devices, designed for working with binary variables. On this basis, one solution to the 1dBPP is provided by means of $b$ binary arrays of size $n$ ($L= \{x_1^{(1)},\dots,x_n^{(1)}\},\{x_1^{(2)},\dots,x_n^{(2)}\}, \dots, \{x_1^{(b)},\dots,x_n^{(b)}\}$). Let us assume a possible instance composed by 10 items ($n=10$). By using a binary representation, one feasible solution comprising two bins ($b=2$) might be:
\[B_1=[0,1,1,0,0,0,1,1,0,0]\]
\[B_2=[1,0,0,1,1,1,0,0,1,1]\]
meaning that four items ($w_2$, $w_3$, $w_7$ and $w_8$) are placed in the first bin, while the remaining six assigned to the second one, hence being the intersection $\emptyset$. This solution can also be encoded as ($L= \{x_1^{(1)}=0,x_2^{(1)}=1,x_3^{(1)}=1,x_4^{(1)}=0,x_{5}^{(1)}=0,x_{6}^{(1)}=0,x_{7}^{(1)}=1,x_{8}^{(1)}=1,x_{9}^{(1)}=0,x_{10}^{(1)}=0\}, \{x_{1}^{(2)}=1,x_{2}^{(2)}=0,x_{3}^{(2)}=0,x_{4}^{(2)}=1,x_{5}^{(2)}=1,x_{6}^{(2)}=1,x_{7}^{(2)}=0,x_{8}^{(2)}=0,x_{9}^{(2)}=1,x_{10}^{(2)}=1\}$). An additional, yet worse, solution for this instance could be:
\[B_1=[0,1,1,0,0,0,1,1,0,0]\]
\[B_2=[1,0,0,0,0,1,0,0,0,1]\]
\[B_3=[0,0,0,1,1,0,0,0,1,0]\]
where three bins are used for storing the complete set of items. As can be deduced, for these solutions to be feasible, each item $w_i\in\mathcal{W}$ should be placed in one, and only one, bin $B_j$, and the capacity $C$ should not be exceeded by any $B_j$.

\section{Proposed Quantum based Heuristic for solving the BPP}\label{sec:solver}

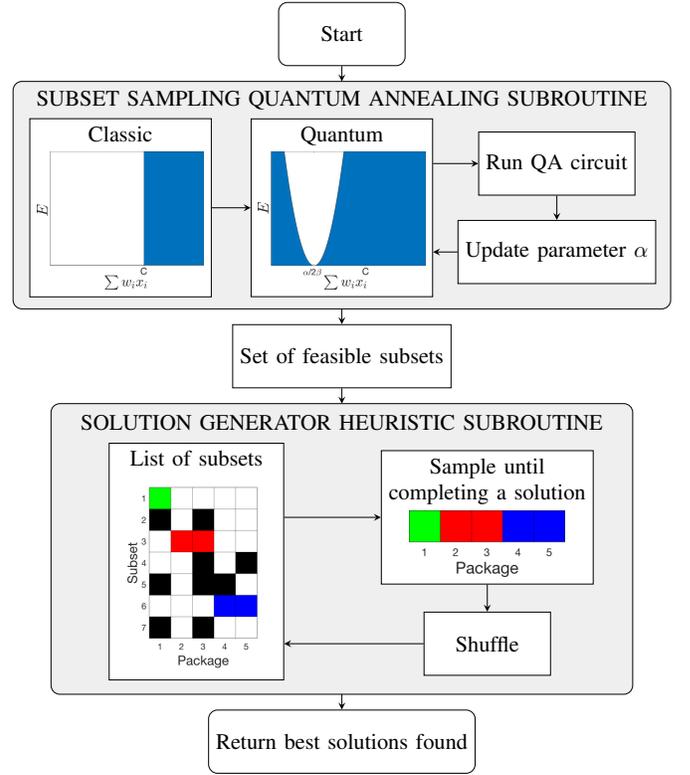
\begin{figure}
\centering
\resizebox{\linewidth}{!}{
\begin{tikzpicture}[node distance=1.5cm]
    \node (start) [startstop] at (0,-0.25) {Start};
    \draw[-stealth] (start) -- (0,-1);
    
    \draw [fill=lightgray!25!, rounded corners] (-5.2,-4.6) rectangle (5.2,-1);
    \node (Qtitle) at (0,-1.3) {SUBSET SAMPLING QUANTUM ANNEALING SUBROUTINE};    
    \node (QclassH) [process, fill=white] at (-3.5,-3) {\shortstack{\phantom{Q}Classic\phantom{Q}\\\includegraphics[width=75pt]{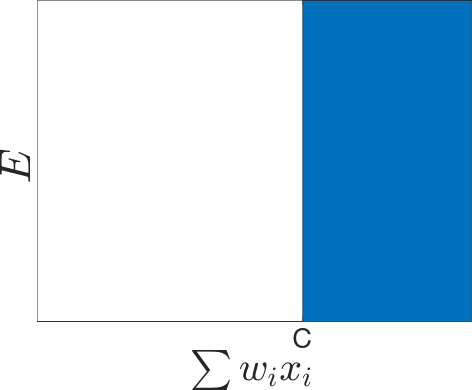}}};
    \node (Qsample) [process, fill=white] at (3.4,-2.3) {\shortstack{Run QA circuit}}; 
    \node (Qchange) [process, fill=white] at (3.4,-3.7) {Update parameter $\alpha$};
    \node (QquantH) [process, fill=white] at (0,-3) {\shortstack{Quantum\\\includegraphics[width=75pt]{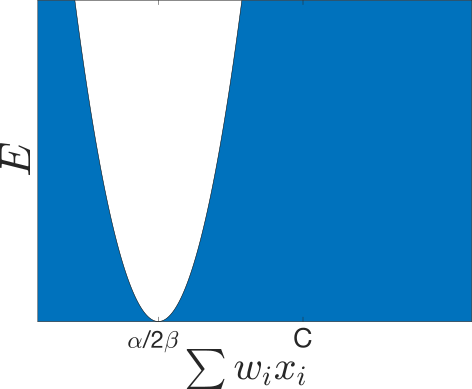}}};
    \draw[-stealth] (Qsample.south) -- (Qchange.north);    
    \draw[-stealth] (QquantH.east|-Qsample.west) -- (Qsample.west);
    \draw[-stealth] (QclassH.east) -- (QquantH.west);
    \draw[-stealth] (Qchange.west) -- (QquantH.east|-Qchange.west);
        
    \node (feasible) [process] at (0,-5.35) {Set of feasible subsets};
    \draw[-stealth] (0,-4.6) -- (feasible.north);
    \draw[-stealth] (feasible.south) -- (0,-6.1);
    
    \draw [fill=lightgray!25!, rounded corners] (-4.6,-10.7) rectangle (4.6,-6.1);
    \node (Qtitle) at (0,-6.4) {SOLUTION GENERATOR HEURISTIC SUBROUTINE};
    \node (Csubsets) [process, fill=white] at (-2.3,-8.6) {\shortstack{List of subsets\\\includegraphics[trim=0 0 30 0,clip,height=90pt]{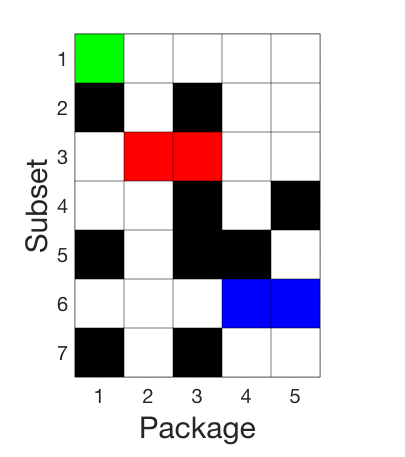}}};
    \node (Csolution) [process, fill=white] at (2.3,-7.9) {\shortstack{Sample until\\completing a solution\\\includegraphics[width=70pt]{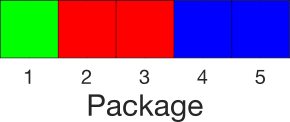}}};
    \node (Cshuffle) [process, fill=white] at (2.3,-9.9) {Shuffle};
    \draw[-stealth] (Csubsets.east|-Csolution.west) -- (Csolution.west);
    \draw[-stealth] (Csolution.south) -- (Cshuffle.north);
    \draw[-stealth] (Cshuffle.west) -- (Csubsets.east|-Cshuffle.west);
    
    \node (end) [startstop] at (0,-11.45) {Return best solutions found};
    \draw[-stealth] (0,-10.7) -- (end.north);
\end{tikzpicture}}
\caption{Diagram of the full H-BPP algorithm. The algorithm is divided into two subroutines. The first one is a quantum annealing algorithm to obtain a set of feasible configuration for single containers. The second one takes the set provided by the quantum subroutine to generate possible solutions. The solution yielded by the algorithm is a list with the best results found.}
\label{fig:diagram}
\end{figure}

In this section, we propose a hybrid quantum-classical algorithm for solving 1dBPP, that is, the H-BPP. The algorithm is divided into two main subroutines. In the first one, a quantum annealing algorithm is executed multiple times to sample feasible configurations of items in a single bin. When this first subroutine ends, a classical subroutine is run, whose objective is to optimally combine the previous list of feasible subsets, considering that each item is assigned to a feasible bin. As a result of this second subroutine, the algorithm returns the best solutions found, this is, the solutions requiring the minimum amount of containers to accommodate all items. A diagram with the overview of H-BPP is shown in Fig. \ref{fig:diagram}. In the rest of the section, we explain these two subroutines in detail.

\subsection{Quantum annealing algorithm for the subset sampling subroutine}\label{sec:QCSub}

The objective of the subset sampling subroutine is to obtain a set of feasible subsets, which effectively reduces the search space for the classical solution construction algorithm. On this regard, we define a feasible subset as a set of items $B_j\in\{1,\dots,n\}$ such that they fulfill the constraint from Equation \ref{1dBPP_const1} for a single container. The process of sampling feasible subsets will be performed with a quantum annealing algorithm.

In a nutshell, quantum annealing is a heuristic quantum algorithm for finding the ground state of a Hamiltonian. By means of an adiabatic evolution, the quantum state evolves from an easily preparable ground state of an initial Hamiltonian $H_0$ to the ground state of the problem Hamiltonian $H_\text{P}$, which encodes the information of the optimization problem. In general, the time-dependent Hamiltonian is given by
\begin{equation}
    H(t)=(1-\Lambda(t))H_0+\Lambda(t)H_\text{P},
\end{equation}
where $\Lambda(t)$ is the mixing function with the boundary conditions $\Lambda(0)=0$, $\Lambda(\tau)=1$, and $\tau$ is the annealing time. For simplicity, and without loss of generality, we choose this function to be a linear function $\Lambda(t)=t$, although it can be designed to verify that the adiabaticity condition is fulfilled \cite{Amin2009AdiabaticCondition}. 

Furthermore, to solve the subset sampling problem, ideally we seek the ground states of a stepwise Hamiltonian
\begin{equation}\label{eq:idealH}
    H_\text{ideal}=\begin{cases}
    \infty & \text{if} \sum_i w_ix_i>C,\\
    0 & \text{otherwise},
    \end{cases}
\end{equation}
where we have slightly modified the notation used in the previous section. Now $x_i=1$ if the $i$-th item belongs to the subset, and $x_i=0$ otherwise. To encode the classical information into the quantum system, we do a one-to-one assignment of each binary variable to a qubit, where we represent the state in the computational basis $x_i=1\rightarrow\ket{1}_i$ and $x_i=0\rightarrow\ket{0}_i$. 

\begin{figure}
    \centering
    \includegraphics[width=\linewidth]{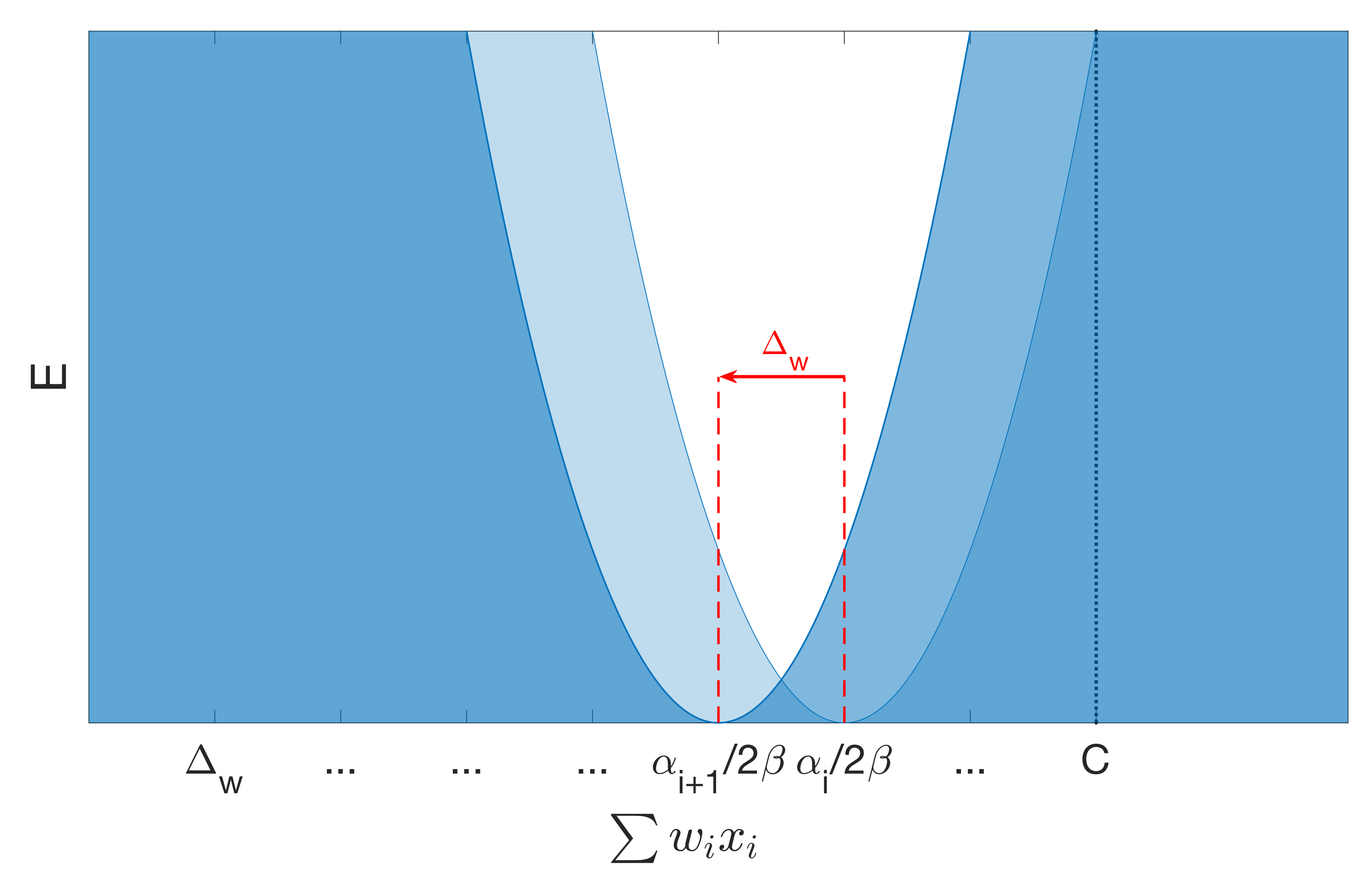}
    \caption{Representation of the sampling strategy used in the 1dBPP problem. The $x$ axis ticks shows the ideal weights of the subsets we are sampling in each step, while the $y$ axis shows the energy of the eigenvalues. If one implements the strategy for not measuring the same eigenvector twice, we would see Dirac delta functions that represent the penalty factors added to the measured states.}
    \label{fig:samplingStrategy}
\end{figure}

Additionally, we cannot generate stepwise Hamiltonians in a real system, so we have to find an implementable model for Equation \ref{eq:idealH}. For this, we employ the following quadratic Hamiltonian
\begin{equation}
\begin{split}
    H_\text{P}&=\alpha\left(\sum_iw_ix_i-C\right)+\beta\left(\sum_iw_ix_i-C\right)^2\\
    &=-\sum_iw_i\left(\frac{\alpha}{2}+\beta\epsilon_w\right)\sigma_i+\sum_{i<j}\frac{\beta w_{i}w_{j}}{2}\sigma_i^z\sigma_j^z,
\end{split}    
\end{equation}
where $\epsilon_w\equiv\sum_iw_i/2-C$ and $\sigma_i^\mu$ is the Pauli $\mu$ operator acting on qubit $i$. $H_\text{P}$ has two-body interactions between all pairs of qubits, i.e. it is an all-to-all (ATA) Hamiltonian. This Hamiltonian can be understood as a quadratic Hamiltonian in which the (possibly degenerated) ground state corresponds to configurations for which $\sum_iw_ix_i=C-\alpha/2\beta$. The two parameters $\alpha,\beta$ allow us to tweak the sum of the weights of the packages, and the energy penalty added to the states not fulfilling the condition. This way, we can sample the allowed configurations by setting $\alpha=2\beta(C-k\Delta_w)$, where $\Delta_w$ is the minimum weight difference between two different configurations and $k=1,...,C/\Delta_w$. Since we cannot know how many runs for the different values of $\alpha$ are required for sampling all possible configurations, we set a total number of runs. Then, we will repeat the annealing algorithm for each value of $\alpha$ and save all the results obtained as shown in Algorithm \ref{alg:QAforSubsetSampling}. A visual representation of the subroutine sampling strategy is displayed in Figure \ref{fig:samplingStrategy}.

\begin{algorithm}
\caption{Quantum annealing subroutine for the subset sampling}
\label{alg:QAforSubsetSampling}
\KwData{$w_i$, $C$, $\beta$, \#runs}
\For{k = 1,...,$C/\Delta_w$}{
    Update Hamiltonian parameter, $\alpha$ = $2\beta(C-k\Delta_w)$\;
    \For{i = 1,...,$\text{\#runs}\ C/\Delta_w$}{
        Run the quantum annealing algorithm, $B_j$\;
        Add the result to the list of subsets, $\mathcal{F}\leftarrow B_j$\;
    }
}
\KwResult{The list of all sampled subsets. $\mathcal{F}$}
\end{algorithm}

Once implemented, we have to run the quantum annealing algorithm. For this, we have two distinct alternatives. On the one hand, we can use a quantum annealer in which we have the resources to implement the evolution under our problem Hamiltonian. On the other hand, we can simulate the digitized evolution, i.e. approximating the time dependant evolution with $n_T$ time-independent evolution steps \cite{Barends2016digitized}, with a DQC or DAQC implementation. With the currently available quantum hardware capabilities and qubit connectivity topology, we can not directly simulate the evolution under $H_\text{P}$. Thus, we have to resort to methods that simulate an ATA Hamiltonian with limited resources, for example, using the decomposition shown in \cite{asier2020enhancedconnectivity}. 

\subsection{Solution Generator Heuristic} \label{sec:ClassicSub}

This section briefly describes the heuristic procedure used to solve the 1dBPP by using as input the set of feasible subsets $\mathcal{F}$ computed in the previous subroutine. Specifically, the proposed scheme aims at seeking, among the elements in $\mathcal{F}$, an optimal combination that packages all items in $\mathcal{W}$. In this case, the search strategy has the objective of finding the configuration with the smallest number of bins. It is important to highlight that this algorithm has been designed to be run in classical devices.

\begin{algorithm}
\caption{Classical heuristic for finding the best combination of bins}
\label{alg:heuristic}
    \KwData{$\mathcal{F}, \mathcal{W}, \var{MaxIter}$}
    \tcc{\var{Initialize solution set}}
    $\mathcal{S} \gets \emptyset$\;
    \For{$c=1,\ldots,\var{MaxIter}$}{
        \tcc{Initialize number of bins}
        $b\gets 0$\;
        \tcc{Initialize the set of assigned items}
        $\mathcal{A} \gets \emptyset$\;
        Shuffle elements is $\mathcal{F}$\;
        \For{$\var{j}=1,\ldots,|\mathcal{F}|$}{
            \If{$\left(B_j\cap\mathcal{A}\right)=\emptyset,\ B_j\in\mathcal{F}$}{
                Include items in $B_j$ in $\mathcal{A}$\;
                \tcc{Update number of bins}
                $b \gets b + 1$\;
            }
            \If{$\mathcal{W} = \mathcal{A}$}{
                $\mathcal{S} \gets \mathcal{S} \cup \{b\}$\;
            }
        }
    }
    \KwResult{$\min \mathcal{S}$}
\end{algorithm}

The pseudocode of the proposed heuristic is given in Algorithm \ref{alg:heuristic}. As aforementioned the procedure gets the set of feasible subsets $\mathcal{F}$ and creates a solution (lines 6-11) by looking for bins (lines 7-9) that might package all the items in $\mathcal{W}$ (line 10). Note that this search is made sequentially in $\mathcal{F}$, so this strategy cannot guarantee the optimal solution if it is computed just once. To improve the search, the algorithm repeats this procedure $\var{MaxItems}$ times (line 2), saving the computed solution in a temporary variable $\mathcal{S}$. Finally, the procedure outputs the solution with the smallest number of bins stored in $\mathcal{S}$.

\section{Experimental Setup, Results and Discussion}\label{sec:exp}

This section is devoted to showcasing the whole experimentation conducted in this study. To this end, this section have been divided into two different subsections, the first aiming at describing the experimental setup (Section \ref{sec:setup}), and the second focused on showing and discussing the main results obtained by our H-BPP (Section \ref{sec:results}).

\subsection{Experimental Setup}\label{sec:setup}

As pointed in the introduction, the benchmark comprises the resolution of 18 different synthetic instances. With the main intention of building an unbiased and heterogeneous benchmark, we have developed an instance generator script. This script takes three different values as configuration parameters: \textit{i}) the number of items ($n$), \textit{ii}) the bin capacity ($C$), and \textit{iii}) the random distribution of item weights that compose the instance. In reference to the number of items, two different values have been considered: \texttt{10} and \texttt{12}. These sizes are sufficiently large for finding meaningful outcomes, regarding the limited resources of the current QC devices. For the maximum capacity, \texttt{100}, \texttt{120} and \texttt{150} values have been deemed. As distribution functions, three configuration have been employed: a single Gaussian centered in $C$/2, a 2-component Gaussian mixture model with centers established in $C/3$ and $2C/3$, and a uniform distribution between 0 and $C$. We summarize the characteristics of the whole benchmark in Table \ref{tab:instances}. Finally, in order to assure the replicability of this study, all the algorithms, instances, and data used in this work are publicly available\footnote{\url{https://bitbucket.org/mikel_gda/bpp}}. 

\begin{table*}[ht!]
	\centering
	\caption{Description of the 18 1dBPP instances generated for measuring the quality of the developed H-BPP algorithm}	
	\resizebox{2.0\columnwidth}{!}{
		\begin{tabular}{ccccc|ccccc}
			\toprule
			Instance & \# of items & min-max weight & capacity & Distribution & Instance & \# of items & min-max weight & capacity & Distribution\\
			\midrule
			\texttt{10\_100\_2G} & 10 & 30-80 & 100 & Double Gaussian & \texttt{12\_100\_1G} & 12 & 10-70 & 100 & Single Gaussian \\
			\texttt{10\_120\_2G} & 10 & 30-80 & 120 & Double Gaussian & \texttt{12\_120\_1G} & 12 & 20-100 & 120 & Single Gaussian \\
			\texttt{10\_150\_2G} & 10 & 40-100 & 150 & Double Gaussian & \texttt{12\_150\_1G} & 12 & 40-120 & 150 & Single Gaussian \\
			\texttt{12\_100\_2G} & 12 & 10-90 & 100 & Double Gaussian & \texttt{10\_100\_U} & 10 & 10-70 & 100 & Uniform \\
			\texttt{12\_120\_2G} & 12 & 30-110 & 120 & Double Gaussian & \texttt{10\_120\_U} & 10 & 10-100 & 120 & Uniform \\
			\texttt{12\_150\_2G} & 12 & 30-120 & 150 & Double Gaussian & \texttt{10\_150\_U} & 10 & 20-140 & 150 & Uniform \\
			\texttt{10\_100\_1G} & 10 & 20-80 & 100 & Single Gaussian & \texttt{12\_100\_U} & 12 & 10-60 & 100 & Uniform \\
			\texttt{10\_120\_1G} & 10 & 20-70 & 120 & Single Gaussian & \texttt{12\_120\_U} & 12 & 10-110 & 120 & Uniform \\
			\texttt{10\_150\_1G} & 10 & 30-120 & 150 & Single Gaussian & \texttt{12\_150\_U} & 12 & 20-130 & 150 & Uniform \\
			\bottomrule
		\end{tabular}
	}
	\label{tab:instances}
\end{table*}

As mentioned in the previous paragraph, the size of the instances has been chosen considering the QC hardware limitations, but also the computational restrictions of the brute-force algorithm used for measuring the quality of our developed H-BPP. On this regard, as a preliminary study, our main objective is to compare the outcomes obtained by our implemented method with the optimal results of each instance. For carrying out this task, we have developed a brute-force algorithm which exhaustively seeks for the minimum number of bins needed for storing all the items of each instance.

The procedure followed by the brute-force algorithm is the following one. First, the technique checks if all items of the instance fit into a single bin (note that none of these cases have been contemplated in our benchmark). Otherwise, it starts an iterative process using a starting exploratory value of $b=2$, and it evaluates all possible feasible combinations for item allocation in $b$ number of bins. In case all items cannot be inserted into $b$ bins due to capacity restrictions, $b$ value is increased by 1. This iterative procedure is executed until the algorithm finds a solution in which all items fit into $b$ bins. After finding the optimal value of $b$, and in the interest of exhaustive experimentation, the algorithm continues the execution in order to find each and every possible item assignment combination considering those $b$ bins. In other words, the brute-force algorithm finds all the optimal solutions, if more than one exists. It is noteworthy that we define two $L$ and $L'$ solutions as different if the arrangement of items $\{x_1^{(1)},\dots,x_n^{(1)}\}, \dots, \{x_1^{(b)},\dots,x_n^{(b)}\}$ differs on its composition, and not in the artificial order promoted by the bin code $j$.

Lastly, the circuit implementing the digitized quantum annealing algorithm has been simulated with a classical computer. As a first approximation to the problem, we have employed an ideal setup, in which we have assumed we have an ATA connectivity and noiseless gates. For the initial Hamiltonian, we used $H_0=\sum_jh_0\sigma_j^x$, where $h_0>0$ is a tweakable parameter. A more rigorous analysis would require to introduce the effect of noise and the extra steps required to simulate the evolution in a hardware with limited resources. For this work, we run the algorithm with the following set of parameters: \#runs=1000, $\beta=min_j(w_j)/5$, $\tau=10^{-14}s$, $n_T=500$, and $h_0=10$. We have selected these parameters based on previous experience solving similar problems and adjusting them to obtain the maximum fidelity within a reasonable range of values \cite{Mikel2022DQAforLND}.

\subsection{Results and Discussion}\label{sec:results}

Table \ref{tab:results} summarizes the outcomes reached by H-BPP for all instances. Each entry on the table depicts the following metrics:

\begin{itemize}
    \item \textit{Fitness average ($\overline{b}$)}: this metric shows the average value of $b$ found by the H-BPP algorithm, considering that the lower the value, the better the result.
    \item \textit{Average number of optima ($\overline{\# optima}$)}: considering that each instance counts with more than one optimal solution, we depict on Table \ref{tab:results} the average number of these optima found by our method.
    \item \textit{Average size of the ideal subset in $\mathcal{F}$}: considering one specific instance of the 1dBPP, the ideal subset is composed of all the $B$ bin configurations needed to build the whole set of optima solutions. On this regard, the \textit{average size of the ideal subset $\mathcal{F}$} is the amount of bins in $\mathcal{F}$ which are part of this complete ideal subset. This metric helps to independently assess the H-BPP performance.
\end{itemize}

These results have been computed over 10 independent runs conducted for each instance. At this point, it should be pointed that $\overline{\# optima}$ and the \textit{average size of the ideal subset in $\mathcal{F}$} are calculated using these runs in which the algorithm has obtained the optimal value. Furthermore, we also show in Table \ref{tab:results} the optimal values for each test case on each of the considered metrics. Thus, we provide the optimal solution (\textit{Opt.}) and the total number of optimum configurations ($\# optima$) for each instance, as well as the size of the optimal ideal subset (\textit{size of opt. ideal sub.}). 

\begin{table*}[ht!]
	\centering
	\caption{Results obtained by the H-BPP and their comparison with the optima values.}	
	\resizebox{2.0\columnwidth}{!}{
		\begin{tabular}{c|ccm{0.08\textwidth}|ccm{0.08\textwidth} |c|ccm{0.08\textwidth}|ccm{0.08\textwidth}}
			\toprule
			& \multicolumn{3}{c|}{H-BPP} & \multicolumn{3}{c|}{Optimum values} & & \multicolumn{3}{c|}{H-BPP} & \multicolumn{3}{c}{Optimum values} \\
			Instance & $\overline{b}$ & $\overline{\# optima}$ & Avg. size of ideal sub. & Opt. & $\# optima$ & size of opt. ideal sub. & Instance & $\overline{b}$ & $\overline{\# optima}$ & Avg. size of ideal sub. & Opt. & $\# optima$ & size of opt. ideal sub. \\
			\midrule
			\texttt{10\_100\_2G} & 6,0 & 20,1 & 22,7 & 6 & 36 & 26 & \texttt{12\_100\_1G} & 6,0 & 103,4 & 61,6 & 6 & 5376 & 110 \\
			\texttt{10\_120\_2G} & 6,0 & 185,3 & 49,8 & 5 & 284 & 59 & \texttt{12\_120\_1G} & 7,6 & 10,2 & 23,2 & 7 & 260 & 38 \\
			\texttt{10\_150\_2G} & 5,0 & 8,0 & 15,5 & 5 & 24 & 20 & \texttt{12\_150\_1G} & 7,3 & 5,8 & 19,8 & 7 & 162 & 34 \\
			\texttt{12\_100\_2G} & 5,4 & 2,1 & 12,1 & 5 & 36 & 31 & \texttt{10\_100\_U} & 4,0 & 688,7 & 113,4 & 4 & 1574 & 111 \\
			\texttt{12\_120\_2G} & 8,4 & 14,1 & 18,8 & 8 & 180 & 29 & \texttt{10\_120\_U} & 6,0 & 140,9 & 48,9 & 6 & 285 & 59 \\
			\texttt{12\_150\_2G} & 6,9 & 1,0 & 13,0 & 6 & 30 & 21 & \texttt{10\_150\_U} & 5,0 & 5,1 & 14,7 & 5 & 8 & 17 \\
			\texttt{10\_100\_1G} & 5,0 & 188,1 & 55,4 & 5 & 500 & 72 & \texttt{12\_100\_U} & 6,0 & 370,7 & 58,3 & 6 & 3132 & 95 \\
			\texttt{10\_120\_1G} & 4,0 & 126,2 & 60,7 & 4 & 337 & 79 & \texttt{12\_120\_U} & 7,0 & 59,9 & 46,1 & 7 & 1230 & 92 \\
			\texttt{10\_150\_1G} & 6,0 & 15,9 & 20,3 & 6 & 42 & 25 & \texttt{12\_150\_U} & 7,3 & 10,4 & 29,2 & 7 & 224 & 54 \\
			\bottomrule
		\end{tabular}
	}
	\label{tab:results}
\end{table*}
Several remarks can be drawn from these outcomes. The first one is to highlight the performance of the proposed method on instances composed by 10 items, in which the H-BPP reaches the optimal results in the $100\%$ of the cases. This trend is maintained regardless the total number of optima or the size of the optimal ideal subset. This has been said assuming that an instance with less number of optima or a reduced optimal ideal subset is harder to be optimized. On this regard, the specific performance demonstrated by H-BPP in the instance \texttt{10\_120\_U} is specially noteworthy. This instance counts with 8 optimum values and a size of optimal ideal subset of 17, making this case the most complex 10-item instance to solve. In any case, H-BPP obtains not only the optimum in all the 10 independent runs (which is the main objective of the method), but a remarkable amount of optima and a significant percentage of the ideal subset. 

Regarding the instances composed by 12 items, despite some results are slightly degraded, the overall performance of the method is still meritorious considering the current capacities of QC resources and that the input parameters of the quantum annealing subroutine have not been optimized in this preliminary study. In this way, there are three specific 12-item instances in which the H-BPP gets the optimal results on each of the 10 independent executions(\texttt{12\_100\_1G}, \texttt{12\_100\_U} and \texttt{12\_120\_U}). If we further analyze the composition of these cases, it can be seen how they count with a high number of optima. This situation makes easier the optimal resolution of these instances. In any case, it can be observed how the H-BPP obtains a significant fraction of the optima in more complex examples, such as \texttt{12\_100\_2G} or \texttt{12\_150\_1G}.

Moreover, we have conducted an additional analysis in Table \ref{tab:noOpt}, in which we compute the \textit{average size of the ideal subset in $\mathcal{F}$} obtained by the H-BPP for the runs in which the optimal solution has not been reached. Our main objective with this is to better understand the behaviour of the method in these specific cases. In this sense, and conducting a deeper analysis of these results, it can be seen in instances such as \texttt{12\_150\_2G}, \texttt{12\_120\_1G} or \texttt{12\_150\_U}, that attaining more than the $50\%$ of the ideal subsets does not guarantee the obtention of any optimal solution. This situation unveils the complexity of some instances, in which the weight of finding some concrete subsets is crucial for reaching the optimum of the problem. In other words, some specific examples are more restrictive than others, since few limited subsets are compulsory needed for obtaining any optima. Not finding these subsets in the quantum annealing algorithm for the subset sampling subroutine (Section \ref{sec:QCSub}) leads to the impossibility of the method to reach the optimal solution. 

\begin{table}[ht!]
	\centering
	\caption{Average size of ideal subset found by H-BPP in those runs that the optimal solution is not found.}	
	\resizebox{\columnwidth}{!}{
		\begin{tabular}{cc|cc}
			\toprule
			Instance & Avg. size of ideal sub. & Instance & Avg. size of ideal sub.\\
			\midrule
			\texttt{12\_100\_2G} & 10,2 (out of 31) & \texttt{12\_120\_1G} & 19,1 (out of 38)\\
			\texttt{12\_120\_2G} & 15,2 (out of 29) & \texttt{12\_150\_1G} & 17,0 (out of 34)\\
			\texttt{12\_150\_2G} & 11,5 (out of 21) & \texttt{12\_150\_U} &  28,3 (out of 54)\\
			\bottomrule
		\end{tabular}
	}
	\label{tab:noOpt}
\end{table}

\section{Conclusions and Future Work}\label{sec:conc}

In this work, we present a new hybrid quantum-classical algorithm for solving 1dBPP. The idea that we have introduced is to use a quantum annealing algorithm to reduce the space of possible solutions. This is performed by searching for feasible configurations of items fitting in a single bin. Within this reduced search space, the algorithm implements a classical heuristic for generating the configuration minimizing the number of containers. As a proof of concept, we have compared the accuracy of our results against a classical brute-force algorithm. Employing a benchmark composed by 18 randomly generated instances with different number of packages and weight distributions, we have found that our algorithm is capable of reliably finding the optimal solution for the smaller instances. For the instances with a higher number of packages, the performance of the algorithm slightly degrades. However, the reduction of the search space and the expected resource consumption of the quantum annealing algorithm in terms of the qubit number, allows us to find approximate solutions for larger instances, while the brute-force algorithm can not find the solution in a reasonable time. 

Despite of the simplicity of the problem solved in this work, this paves the way for addressing more complex BPP problem variants (such as the two-dimensional or three-dimensional BPPs) using hybrid quantum-classical algorithms. There is room for improvement regarding the quantum annealing subroutine, and the selection of its hyper-parameters. As a future work, we have planned to improve the classical module of the H-BPP, in order to increase the probability of finding the subset of packages found in the optimal configurations. In this regard, we need to perform further analysis of the algorithm to prove the possible advantage of using a complete quantum algorithm for the whole process.

\section*{Acknowledgments}

The research leading to this paper has received funding from the QUANTEK project (ELKARTEK program from the Basque Government, expedient no. KK-2021/00070). Additionally, MS acknowledges support from Spanish Ram\'on y Cajal Grant RYC-2020-030503-I, the projects QMiCS (820505) and OpenSuperQ (820363) of the EU Flagship on Quantum Technologies, from the EU FET Open project Quromorphic (828826) and EPIQUS (899368), as well as from IQM Quantum Computers project "Generating quantum algorithms and quantum processor optimization". MGdA acknowledges support from the UPV/EHU and TECNALIA 2021 PIF contract call.

\section*{Note}

This article has been submitted to the Genetic and Evolutionary Computation Conference 2022 (GECCO 2022).

\bibliographystyle{IEEEtran}
\bibliography{IEEEexample}

\end{document}